\documentclass[1, 12pt]{elsarticle}
\usepackage{amsmath}
\usepackage[toc,page]{appendix}
\usepackage{amssymb}       
\usepackage{amsthm}       
\usepackage{natbib}        
\usepackage{graphicx}
\usepackage{graphics}
\usepackage{booktabs}
\usepackage{subfigure}      
\usepackage{color}
\usepackage{gensymb} 
\graphicspath{{Figures/}}
\usepackage[version=4]{mhchem} 
\usepackage{upgreek}

\setcitestyle{square, comma, numbers,sort&compress}
\makeatletter
\def\@xfootnote[#1]{%
  \protected@xdef\@thefnmark{#1}%
  \@footnotemark\@footnotetext}
\makeatother

\journal{Physical Chemistry Chemical Physics, accepted for publication}

\begin{document}

\begin{frontmatter}

\title{Understanding the effect of mechanical strains on the catalytic activity
of transition metals}

  \author[imdea,ucm]{Carmen Mart\'inez-Alonso}
  \author[upm]{Jos\'e Manuel Guevara-Vela}
  \author[imdea,upm]{Javier LLorca\corref{cor1}}
  \cortext[cor1]{To whom correspondence should be addressed:
javier.llorca@upm.es , javier.llorca@imdea.org}

  \address[imdea]{IMDEA Materials Institute, C/Eric Kandel 2, 28906 - Getafe,
Madrid, Spain.}
  \address[ucm]{Department of Inorganic Chemistry, Complutense University of
Madrid, 28040 Madrid, Spain.}
  \address[upm]{Department of Materials Science, Polytechnic University of
Madrid, E. T. S. de Ingenieros de Caminos, 28040 Madrid, Spain.}

 \begin{abstract}

The effect of elastic strains on the catalytic activity for the hydrogen
evolution reaction (HER) and the oxygen reduction reaction (ORR) was analyzed
on thirteen  late transition metals:  eight (111) surfaces of fcc metals (Ni, Cu, Pd, Ag, Pt, Au, Rh, Ir) and five (0001) surfaces of hcp metals (Co, Zn, Cd, Ru, and Os).
The corresponding adsorption energies for the different intermediate reactions
up to strains dictated by the mechanical stability limits  were previously
obtained by means of density functional theory calculations. It was found that
the elastic strains can be used to tune the catalytic activity of different
metals by reducing the energy barrier of the rate limiting step and even to
reach the cusp of the volcano plot. The largest changes in catalytic activity
with strain for the HER were found in Pt, Au, and Ir while Co and Ni were very
insensitive to this strategy. In the case of the ORR, the catalytic activity of
Au could be enhanced by the application of tensile strains  while that of Cu,
Ni, Pt, Pd, Rh, Co, Ru, and Os was improved by the application of compressive strains.
However, the catalytic activity of Ir was rather insensitive to mechanical
deformations. Elastic strains were able to modify the rate limiting reaction in
Au, Pt, Ag, and Os  and it was possible to achieve the cusp of the volcano plot in
these metals. Final, mechanical instabilities  were attained at small strains
in Zn and Cd, which did not lead to significant changes in the catalytic
activity for HER and ORR. These results provide a framework to systematically
investigate the application of elastic strains in the design of new catalysts.

 \end{abstract}


\end{frontmatter}

\parskip  5pt

\section{Introduction}

The hydrogen economy, which stands as one of the most promising pillars of the
future of clean energy, strongly depends on two key reactions for the
production of hydrogen and the generation of green energy: the hydrogen
evolution reaction (HER) -- whose global process in acidic media is H$^+$ + e$^-$ $\rightarrow$ 0.5H$_2$ --,  and the oxygen reduction reaction
(ORR) -- whose global process in acidic media is O$_2$ + 4(H$^+$ + e$^-$) $\rightarrow$ 2H$_2$O --~\cite{Vesborg2015,Shao2016}. Without an appropriate catalyst, these
processes take place at high temperatures and pressures, which hinders the
widespread use of hydrogen in industrial applications.  So far, the role model
catalyst for the HER and the ORR is platinum, but its high cost and limited
availability drives the research to look for more effective and less expensive
electrocatalysts~\cite{Shao2016,Hansen2021}.  Traditionally, three different
strategies have been used to search new materials that fulfill these criteria:
the introduction of surface
defects~\cite{Fu2019,ZamoraZeledn2021,Mani2008,STRASSER2016166,YANG201772}, the
exploration of different facets~\cite{Li2015,Duan2011}, and
alloying~\cite{Li2018,
SHI2018442,Liu2018,ELDEEB2015893,LIN2015274,Tian2018,YING2014214,CHEN2014380}.

As an alternative, the application of elastic strains to modify the catalytic
properties of materials has been comparatively less explored but there is
recent evidence of promising results in
metals~\cite{Shuttleworth2017,Shuttleworth2016,Shuttleworth20177,Shuttleworth2018,Mollaamin2008,Raman2018,Shen2017,Verga2018,Pati2011}
and other materials~\cite{Grabow2006,Wang2018b}. Indeed, mechanical strains may
have large effects on different physical and chemical properties through the
modification  of the electronic structure. It has been shown that elastic
strains can minimize defect formation in halide
perovskites~\cite{Kim2020,Saidaminov2018}, prevent non-radiative
recombination~\cite{Jones2019}, and even increase device
efficiency~\cite{Liu2021}. The application of elastic strain can also enhance
corrosion~\cite{Shi2020} and oxidation processes~\cite{Pratt2013}. On
transition metals, numerous studies show that mechanical strain can produce
important changes in their
reactivity~\cite{Bhattacharjee2016,Grunze1982,Rao1991,Ruban1997,Cheng1995,Xin2014}.

The adsorption energy of the reactants constitutes one of the best descriptors
in many catalytic processes and  the effect of mechanical strains in the
adsorption energy of different metals has been analyzed in various publications
~\cite{Shuttleworth2017,Shuttleworth2016, Shuttleworth20177,Shuttleworth2018},
including our recent study of the effects of elastic strain on the adsorption
of H, O, and OH onto the surfaces of 11 transition
metals~\cite{MartnezAlonso2021}. However, the influence of each reaction
intermediate should be taken into account to assess the catalytic activity in
processes in where many steps are involved, such as the ORR, and this analysis
is missing for the catalytic activity of transition metals for the HER and the
ORR.   For instance, it is known that the application of tensile (compressive)
strains can decrease (increase) the adsorption energy of H, O, and OH on
transition metal surfaces ~\cite{MartnezAlonso2021,2021} but the influence of
these changes on the catalytic activity has not been explored systematically
for the HER and ORR.

Within this context, the present article analyzes the effect of mechanical
strains onto the catalytic properties of 8 fcc (Ni, Cu, Pd, Ag, Pt, Au, Rh, Ir)
and 5 hcp (Co, Zn, Cd, Ru, Os) transition metals for the HER and the ORR. It was found
that elastic strains modify the catalytic activity for all the studied metals:
compressive strains were found to favor the catalytic properties of Pt, Pd, Rh,
Ni, Ir, Co, Ru, and Os  in the HER and Pt, Pd, Cd, Ir, Zn, Rh, Cu, Ni, Co, Ru, and Os  in the
ORR, while tensile strains improved the catalytic properties of Cu, Au, and Ag
in the HER, and Au and Ag in the ORR.  Nevertheless, metals presented different
sensitivity to the effect of elastic strains on the catalytic activity, which
depended on the rate limiting step. It was found that the optimum strain to
attain the maximum catalytic activity was determined by either the maximum
strain that can be achieved before reaching the mechanical stability limit or
when the rate limiting step changes from one reaction to another. Thus, the
results on this paper rationalize the mechanisms that determine the influence
of mechanical strains on the catalytic activity of transition metals and
provide a framework to explore their effect on other materials.

\section{Methodology}

The variation of the adsorption energy with strain can be used to modulate the
activity of a catalyst following Sabatier$'$s principle~\cite{Sabatier}. It
states that the adsorption energy should be neither too high nor too low for
reactions passing through an adsorbed intermediate. If the energy is too high
(endothermic), adsorption is slow and limits the overall rate, whereas the
catalyst surface becomes poisoned and desorption is limited if the adsorption
energy is too low (exothermic). In terms of hydrogen and oxygen
electrocatalysis, this principle leads to the conclusion that  the free energy
of adsorption should be close to zero at the equilibrium
potential~\cite{Viswanathan2012,Xie2017}. If Sabatier's principle is the only
factor that governs the catalytic process, the plot of the reaction rate versus
the energy of adsorption of the intermediate species leads to a volcano-shaped
curve~\cite{Nrskov2004,Nrskov2005}. Starting from a low, negative (exergonic)
free energy of adsorption, $G_\mathrm{ads}$, the catalytic activity initially
rises with $G_\mathrm{ads}$, leading to the ascending branch of the volcano.
The rate passes through a maximum around $G_\mathrm{ads}$ = 0, and then starts
to decrease as $G_\mathrm{ads}$ becomes more endergonic in the descending
branch of the volcano. The magnitude of $E_\mathrm{ads}$ -- obtained from DFT
calculations -- can be used to determine  the overall rate of the reaction
according to

\begin{equation}
G_\mathrm{ads} = E_\mathrm{ads} + E_\mathrm{ZPE} - T\Delta S
\label{ecgibbs}
\end{equation}

\noindent where $G_\mathrm{ads} $, $E_\mathrm{ads} $, $E_\mathrm{ZPE}$, and
$\Delta S$ stand for the variation of the free and adsorption energies, the
zero point energy and the entropy, respectively, during the adsorption of the
intermediate species, and $T$ is the absolute temperature. This relationship
should be evaluated for each intermediate species that appears in the catalytic
process (only H in the HER reaction and O and OH in the ORR), taking into
account that the process with the highest free energy barrier will limit the
rate of the reaction.

 The metal surfaces of the  thirteen metals were subjected to normal and shear
stresses in the surface plane and the adsorption energies were calculated in
the most favorable adsorption site (FCC for fcc (111) metals and HCP for hcp
(0001) metals). Mixed boundary conditions are imposed to solve the elastic
problem in the DFT calculations. They include imposed strains in the slab plane
and zero stresses on the free surface. The deformation gradient ${\mathbf F}$
applied to the supercell was

\begin{equation}
\begin{array}{ccc}
{\mathbf F}= \begin{pmatrix}
1+\epsilon_1 & \gamma & 0\\
0 & 1+\epsilon_2 & 0\\
0 & 0 & 1
\end{pmatrix}
\end{array}
\end{equation}

\noindent where  $\epsilon_1$ and $\epsilon_2$ stands for the normal strains
along $x$ and $y$ directions while and $\gamma$  for the shear distortion in
the $xy$ plane. Uniaxial deformation was applied when $\epsilon = \epsilon_1$
and $\epsilon_2 = \gamma$ = 0, while $\epsilon = \epsilon_1 = \epsilon_2$ and
$\gamma =0$ for biaxial deformation.

\subsection{Hydrogen Evolution Reaction}

The HER is the cathodic reaction in the process of water splitting and plays a
key role in the production of hydrogen by dissociation of the water molecule.
It is a classic example of a reaction with transference of two electrons
through the Volmer-Heyrovsky or Volmer-Tafel mechanisms~\cite{Jiao2015}. The
kinetics of the reaction is limited by the adsorption of hydrogen on the
surface for acidic solutions. The adsorption process (Volmer) limits the
kinetics if hydrogen binds weakly on the catalyst surface, while desorption
(Heyrovsky/Tafel) is the limiting process otherwise.  Thus, the adsorption
energy of hydrogen can be used to determine the optimum catalyst for the HER
reaction. The global process in acidic media can be represented as

\begin{center} 
\ce{H+ + e- ->1/2H2} 
\end{center} 

\noindent but the mechanism involves first the electrochemical hydrogen adsorption
(Volmer reaction) followed by the electrochemical (Heyrovsky reaction) and/or
chemical (Tafel reaction) hydrogen desorption reactions

\begin{center} 
\ce{H+ + $*$ + e- -> H$^*$}
\end{center} 

\begin{center} 
\ce{H$^*$ + H+ + e- ->$*$ + H2}
\end{center}

\begin{figure}[t!]
  \centering
  \includegraphics[width=0.9\textwidth]{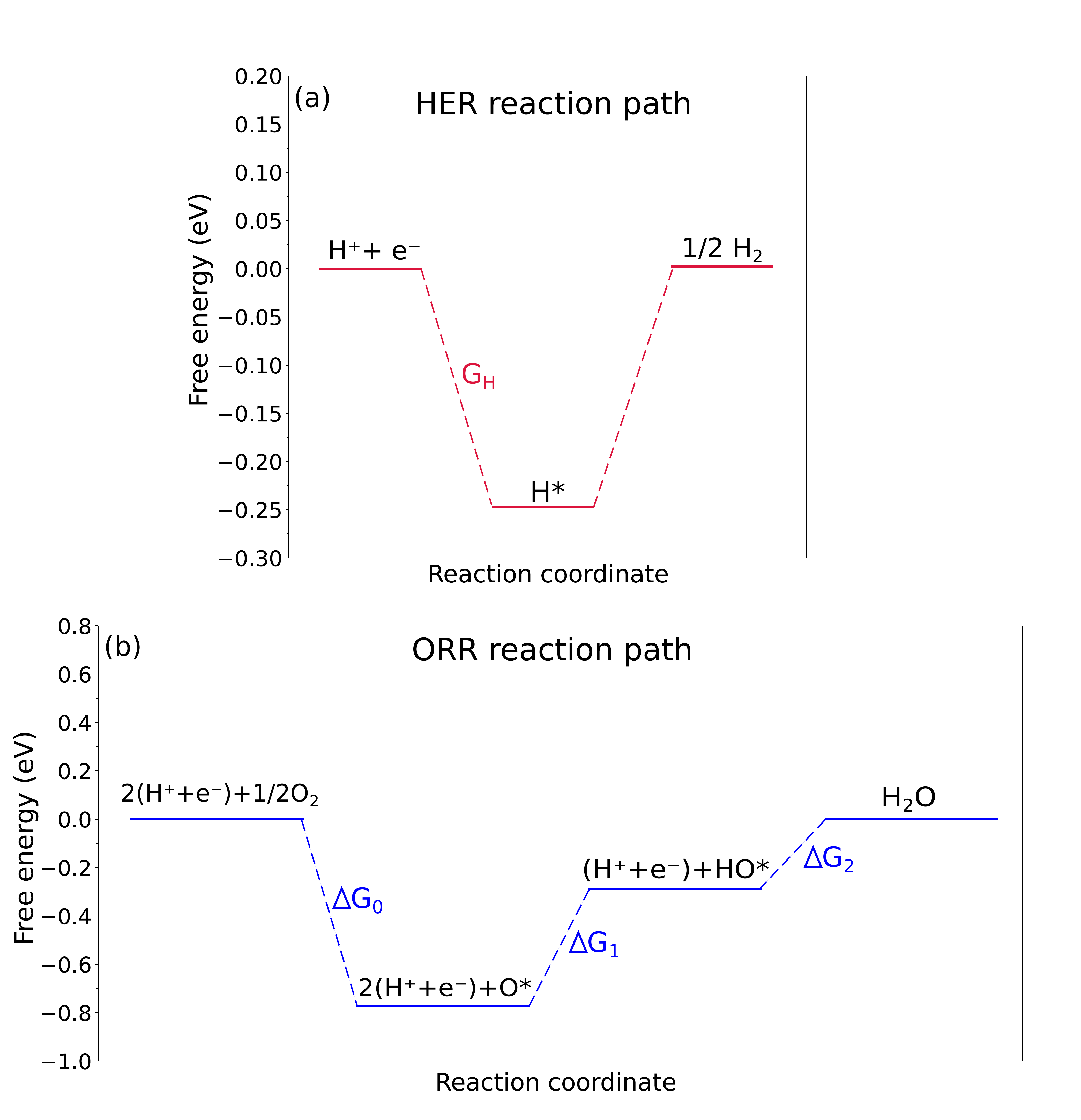}
  \caption{(a) Reaction path for the HER. (b) Reaction path for the ORR (dissociative mechanism). The free energies have been
calculated for \{111\} Pt adsorbed at FCC positions without applied strain.}
  \label{mecanismos}
\end{figure}

\begin{center} 
\ce{2H$^*$  -> 2$*$ + H2}
\end{center}

\noindent where $*$ represents a site on the surface of the catalyst and H$^*$
the hydrogen atom adsorbed. A more detailed diagram of the different steps is shown in Figure 1(a) of the supporting information. The reaction path for the HER
is shown in Figure \ref{mecanismos}(a) and equation \eqref{ecgibbs} can used to compute the free energy associated with
the adsorption process, $G_\mathrm{ads}$, including the differences in zero-point energies between products and reactants, $\Delta E_\mathrm{ZPE}$. These zero-point energies were calculated for the adsorption of H on Cu(111) and on
Pt(111) and their values were 0.04 eV in both cases, in agreement with the one previously reported in
the literature for the adsorption of H on Cu(111)~\cite{Nrskov2005}. Details
about these calculations can be found in section two of the supporting
information. The same value of $\Delta E_{ZPE}$ was used for all the metals
studied below.  

\noindent The vibrational entropy is determined by the contribution of 0.5H$_2$

\begin{align}
\bigtriangleup S = - \frac{1}{2}S^{0}_{H_2}
\end{align}

\noindent where $S^{0}_{H_2}$ = 130.68 J/mol*K ~\cite{entropy}  is the entropy
of H$_2$  in the gas phase at standard conditions (300K and 1 bar).

\noindent Thus, $G_\mathrm{ads}$  at $T$ = 300 K can be expressed as

\begin{align}
G_\mathrm{ads_H} = E_\mathrm{ads_H} + 0.24 \ eV
\label{gibbsH}
\end{align}

\noindent$ G_\mathrm{ads_H}$ depends only on the energy of adsorption of
hydrogen according to equation (\ref{gibbsH}) and the catalytic activity of the
HER can be measured by the exchange current density, i$_0$. This parameter
describes the electron transfer between the electrode (which in our case would
be the catalyst) and the solution and can be calculated following the strategy
developed by  Norskov {\it et al.}~\cite{Nrskov2005}. If $G_\mathrm{ads_H} <$
0, the proton transfer is an exothermic process and the exchange current
density can be expressed as

\begin{align}
i_0 = -e  \hspace{0.05 cm} k_0 \hspace{0.05 cm} \frac{1}{1 + exp \hspace{0.05 cm}(-  G_\mathrm{ads_H} /kT)}
\label{i1}
\end{align}

\noindent where $e$ is the electron charge and  $k_\mathrm{0}$ = 200 s$^{-1}$
site $^{-1}$~\cite{Nrskov2004} is a constant independent of the metal that
includes all the effects related to the reorganization of the solvent during
the transfer of protons to the surface.  $kT$ stand for the Boltzmann constant
and the absolute temperature, respectively. On the contrary, if
$G_\mathrm{ads_H}$ $>$ 0, the proton transfer is  endothermic, and the exchange
current density is given by

\begin{align}
i_0 = -e  \hspace{0.05 cm} k_0 \hspace{0.05 cm} \frac{1}{1 + exp \hspace{0.05 cm}(-  G_\mathrm{ads_H} /kT)} \hspace{0.1 cm} exp \hspace{0.05 cm}(- G_\mathrm{ads_H} /kT)
\label{i2}
\end{align}

\subsection{Oxygen Reduction Reaction}

The ORR is the critical reaction for the generation of  energy from hydrogen in
a fuel cell. Currently, the kinetics of the ORR is limited by the low rate of
the reduction reaction of oxygen in acidic media at the cathode

\begin{center} 
\ce{O2 + 4(H+ + e-) -> 2H2O} 
\end{center} 

\noindent and it has to be enhanced by means of Pt-based catalysts. The high cost  and low
availability of Pt hinders the widespread application of proton exchange membrane
fuel cells (PEMFC) in the transportation sector. The ORR in a PEMFC is carried
out through the transfer of 4e- through either an associative or dissociative
pathway~\cite{PhysRevLett.93.116105}. The catalytic activity is limited by the
electron-proton transfer to O$^*$ or to OH$^*$ if oxygen is strongly adsorbed on the
surface. Otherwise, the catalytic activity is controlled by the electron-proton
transfer to O$^*$ or by the fracture of the O-O bond, depending on the applied
potential~\cite{PhysRevLett.93.116105}. The energies associated with these
processes are controlled by the electronic structure of the catalyst and can be
modified by the application of elastic strains~\cite{MartnezAlonso2021}. 

Two different mechanisms, dissociative or associative, can be used to describe
the ORR. The molecule of O$_2$ dissociates before it is hydrogenated in the
former, whereas hydrogenation happens previously to the dissociation in the
latter. A more elaborate diagram of both mechanisms is shown in Figure 1(b) of the supporting information. The steps of both mechanisms are show below

Dissociative mechanism:

\begin{center} 
\ce{1/2O2 + $*$ -> O$^*$}
\end{center} 

\begin{center} 
\ce{O$^*$ + H+ + e- ->HO$^*$}
\end{center}

\begin{center} 
\ce{HO$^*$ +  H+ + e-  -> H2O + $*$ }
\end{center}

Associative mechanism:

\begin{center} 
\ce{O2 + $*$ -> O2$^*$}
\end{center} 

\begin{center} 
\ce{O2$^*$ + H+ + e- ->HO2$^*$}
\end{center}

\begin{center} 
\ce{HO2$^*$ +  H+ + e-  -> H2O + O$^*$ }
\end{center}

\begin{center} 
\ce{O$^*$ +  H+ + e-  -> HO$^*$}
\end{center}

\begin{center} 
\ce{HO$^*$ +  H+ + e-  -> H2O + $*$ }
\end{center}

Although each mechanism presents different reactions, the reaction rate is independent of the mechanism 
and this analysis will be based on the dissociative mechanism, which is simpler. Two different
adsorbed species have to be considered in this case,
O$^*$ and HO$^*$ (notice the difference with the HER, where the only
intermediate adsorbate is H$^*$). The reaction path for
the dissociative mechanism in depicted in Figure \ref{mecanismos}(b).
Three different free energies, $G_\mathrm{0}$, $G_\mathrm{1}$, and
$G_\mathrm{2}$, that correspond to the three steps of the dissociative
mechanism shown above, are involved in the process. As detailed in the supporting information, following eq.
\eqref{ecgibbs}, they can be expressed as

\begin{align}
G_\mathrm{0}=G_\mathrm{O^*+ H_2} - G_\mathrm{H_2O + *}= E_\mathrm{adsO'} + 0.01\ eV
\label{G0sim}
\end{align}

\begin{align}
G_\mathrm{1}=  G_\mathrm{HO^* + \frac{1}{2}H_2} - G_\mathrm{O^* + H_2} =E_\mathrm{adsOH} -  E_\mathrm{adsO'} -0.26\ eV
\label{G1sim}
\end{align}

\begin{align}
G_\mathrm{2}= G_\mathrm{H_2O + *} - G_\mathrm{HO^* + \frac{1}{2}H_2} =- E_\mathrm{adsOH} +0.25\ eV
\label{G2sim}
\end{align}

\noindent where $E_\mathrm{adsO'}$ is defined below and  it is assumed that $T$ = 300 K. $G_\mathrm{0}$ and $G_\mathrm{2}$ depend only on the
adsorption energy of O and OH, respectively, whereas $G_\mathrm{1}$ depends on
the difference. 

Additionally, the effect of the potential in the cell has to be taken into
account because it is an electron transfer process. The free energy at pH = 0,
 atmospheric pressure of 1 bar, and 300 K of temperature, with
electrode potential corrections, is given by

\begin{align}
G(U)= G + neU
\label{potential}
\end{align}

\noindent where $U$ is the cell potential, $n$ the number of electrons that
flow from the media to the electrode (if the electrons move in the opposite
direction, $n$ must be multiplied by -1), and $e$  the elementary charge
carried by a single electron, -1. At equilibrium, $U$ would be the equilibrium
potential U$_0$=1.23 V \cite{Nrskov2004}. Considering the electrons that flow
in each of the three steps, the corresponding free energies at equilibrium
potential, atmospheric pressure and 300 K are:

\begin{align}
G_\mathrm{0}(U_0)= E_\mathrm{adsO'} + 0.01eV -2U_0 = E_\mathrm{adsO'} -2.45 \ eV
\label{G0pot}
\end{align}

\begin{align}
G_\mathrm{1}(U_0) =E_\mathrm{adsOH} -  E_\mathrm{adsO'} -0.26eV + U_0= E_\mathrm{adsOH} -  E_\mathrm{adsO'} + 0.97 \ eV
\label{G1pot}
\end{align}

\begin{align}
G_\mathrm{2}(U_0)= - E_\mathrm{adsOH} +0.25eV + U_0= - E_\mathrm{adsOH} + 1.48\  eV
\label{G2pot}
\end{align}

The catalytic activity, $A$, can be expressed \cite{Nrskov2004}

\begin{align}
A = kT \hspace{0.1 cm} \log \big[exp \hspace{0.05 cm}(-G(U_0)/kT)\big]
\label{activityorr}
\end{align}

\noindent where  $G(U_0) = \max\{G_0(U_0), G_1(U_0), G_2(U_0)\}$ and
corresponds to the  activation energy of the process that limits the catalytic
reaction.

\subsection{Adsorption energies}

The adsorption energies of H ($E_\mathrm{adsH}$), O ($E_\mathrm{adsO}$), and OH
($E_\mathrm{adsOH}$) on  thirteen  transition metals subjected different elastic
strains were calculated using density functional theory by Martínez-Alonso
{\it et al.}~\cite{MartnezAlonso2021}. The adsorption energy,
$E_\mathrm{ads{O'}}$, takes into account that the adsorbed oxygen comes from
water in the catalytic process from the reaction:

\begin{center} 
\ce{H2O + $*$ -> O$^*$ + H2}
\end{center}
\setlength{\parskip}{0mm}

\noindent and it is expressed as 

\begin{align}
E_\mathrm{adsO'} = E_\mathrm{O^*} + E_\mathrm{H_{2}} - E_\mathrm{H_{2}O} - E_{*} = E_\mathrm{adsO} + E_\mathrm{H_{2}} - E_\mathrm{H_{2}O} + \frac{1}{2} E_\mathrm{O_2}
\end{align}

\noindent where $E_\mathrm{H_{2}}$, $E_\mathrm{O_{2}}$, and $E_\mathrm{H_{2}O}$
account for the total energy of the hydrogen, oxygen, and water molecules in
the gaseous state, respectively.

The metal surfaces of the thirteen metals were subjected to three different types
of strain: biaxial and uniaxial (tension and compression) as well as  shear,
and the adsorption energies at the most favorable adsorption site (FCC for fcc
(111) surfaces and HCP for hcp (0001) surfaces) were determined. The maximum
strains were established taken into account the mechanical stability limits for
each metal and surface obtained from phonon calculations and varied from -5\%
compression to 8\% tension in Ni, Cu, Pd, Ag, Pt, Au, Rh, Ir, Co, Ru, and Os and from
-2\% compression to 8\% tension in Cd and Zn.

\parskip 5pt
\section{Results} 

\subsection{Volcano plots at mechanical equilibrium}

Gerischer~\cite{Gerischer2010} and Parsons ~\cite{Parsons_1958} were the first ones
to find out that certain models for the activity of the HER resulted in a
volcano-like curve. Nonetheless, it was Trasatti ~\cite{TRASATTI1972163}  who
collected experimental data and represented the first volcano plot for the HER.
Since then, many authors have represented the exchange current density as a
function of the free energy  for hydrogen adsorption for the HER and the
catalytic activity against the adsorption energy of O for the ORR
~\cite{Nrskov2005,Quaino2014,Greeley_2006,10.1093/nsr/nwx119}.  These plots
provide a very intuitive and visual representation of the catalytic performance
of transition metals. The peak of the volcano corresponds to the optimum
catalytic properties, the ascending brand corresponds to metals in which the
reactions are limited by the desorption of the products, and the descending
branch represents the processes that are limited by the adsorption of the
reactants. According to Sabatier$'$s principle~\cite{Sabatier}, an equilibrium
between adsorption of the reactants and desorption of the products should be
attained to get the maximum activity. 

\begin{figure}[!]
 \centering
  \includegraphics[width=0.8\textwidth]{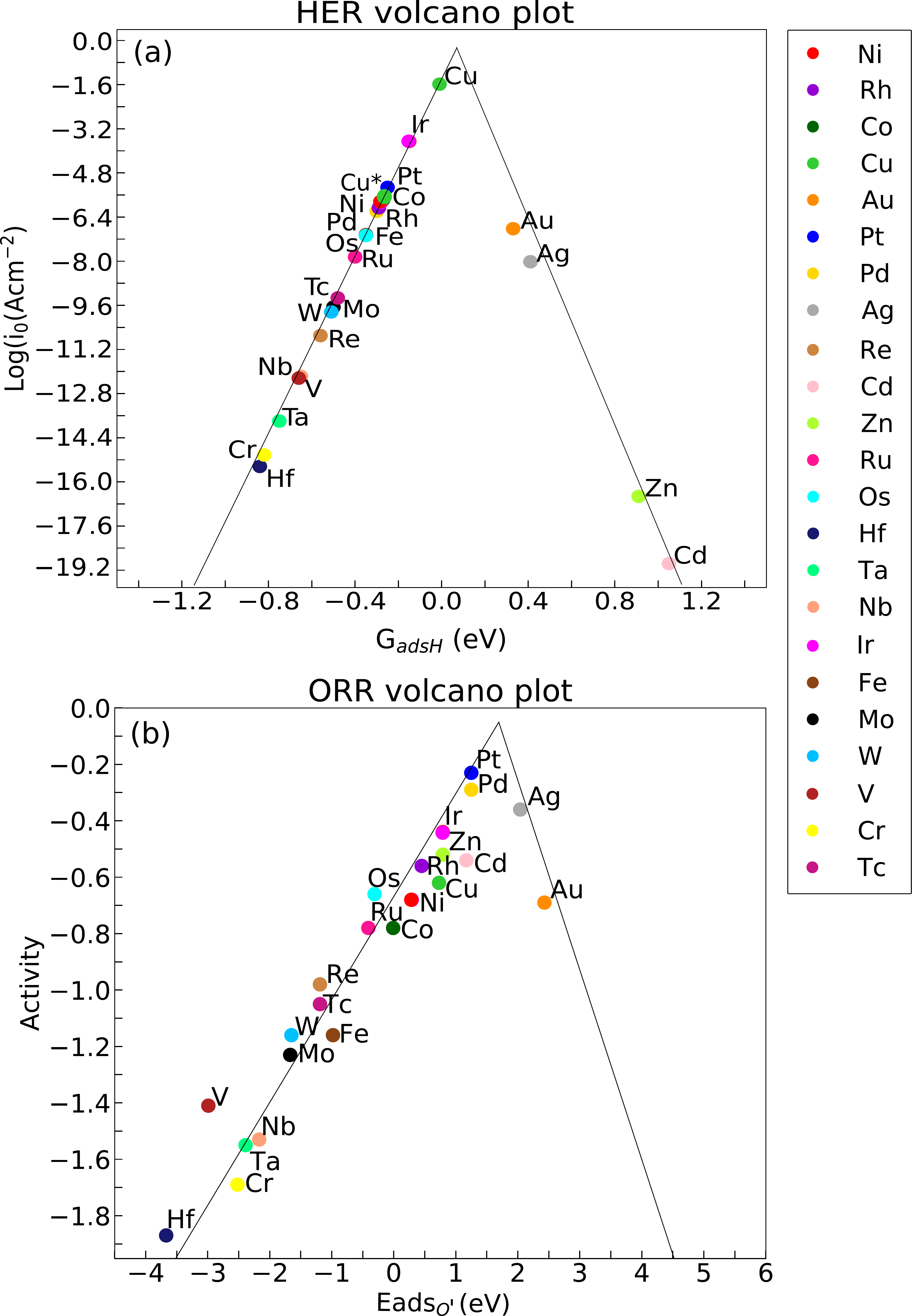}
  \caption{Volcano plots at mechanical equilibrium for (a) the HER as a
function of $G\mathrm{_{adsH}}$ and (b) the ORR as a function of E$_{adsO'}$.  Cu* in the HER volcano plot corresponds to the free energy obtained from the adsorption energy calculated with the DFT functional SCAN~\cite{Sun2015}.}
  \label{volcanoseq}
\end{figure}

The adsorption energy of hydrogen at equilibrium in the most favorable
adsorption site, $E_\mathrm{adsH}$, is given in Table
\ref{gher}~\cite{MartnezAlonso2021}. This information can be used to calculate
the free energy and the corresponding exchange current density according to
eqs. \eqref{i1} and \eqref{i2}, which are also included in Table  \ref{gher}.
Volcano plots at mechanical equilibrium of 23 transition metals for the HER
obtained from the DFT calculations are plotted in Figure \ref{volcanoseq}(a).
 A detailed comparison between the experimental data from the literature and the calculated exchange current densities as a function of free energy of adsorption of H  is shown in section 3 of the supporting information.
These results are in good agreement with the ones found in 
literature~\cite{Nrskov2005,Quaino2014,Greeley_2006,10.1093/nsr/nwx119} with the
exceptions of Cu and Ir, whose experimental activity is lower than that
computed according to Figure \ref{volcanoseq}(a).  In particular, the
$G_\mathrm{adsH}$ calculated for Cu is around 0.2 eV lower than what has been
previously stated, which places Cu on the cusp of the volcano plot. 

\begin{centering}
\begin{table}[!]
\centering
\caption{Adsorption energy of hydrogen at equilibrium in the most favorable
adsorption site, $E_\mathrm{adsH}$, for different transition metals. The free energy of adsorption,  $G_\mathrm{adsH}$, at 300 K and the logarithm of the exchange current density, $\log i_0$, for the HER according to eqs. \eqref{i1} and \eqref{i2} are also included.}
\begin{tabular}{cccc}
\toprule
Metal & $E_\mathrm{adsH}$ (eV) & $G_\mathrm{adsH}$ (eV) & $\log i_0$ (A cm$^{-2}$) \\
\midrule
Pt    & -0.49      & -0.25          &  -5.52          \\
Au    &  0.09      &  0.33          &  -7.06          \\
Cu    & -0.25      & -0.01          &  -1.52          \\
Ag    &  0.17      &  0.41          &  -8.27          \\
Pd    & -0.54      & -0.30          &  -6.40          \\
Ni    & -0.52      & -0.28          &  -6.07          \\
Ir    & -0.39      & -0.15          &  -3.83          \\
Rh    & -0.53      & -0.29          &  -6.27          \\
Cd    &  0.81      &  1.05          & -19.39          \\
Zn    &  0.67      &  0.91          & -16.92          \\
Co    & -0.51      & -0.27          &  -5.97          \\
Nb    & -0.89      & -0.65          & -12.46          \\
Mo    & -0.74      & -0.50          &  -9.92          \\
W     & -0.75      & -0.51          & -10.09          \\
Re    & -0.80      & -0.56          & -10.97          \\
Ru    & -0.64      & -0.40          &  -8.08          \\
Fe    & -0.59      & -0.35          &  -7.26          \\
Os    & -0.59      & -0.35          &  -7.29          \\
Hf    & -1.08      & -0.84          & -15.77          \\
Ta    & -0.99      & -0.75          & -14.11          \\
V     & -0.90      & -0.66          & -12.53          \\
Cr    & -1.06      & -0.82          & -15.35          \\
Tc    & -0.72      & -0.48          &  -9.58          \\
\bottomrule
\label{gher}
\end{tabular}
\end{table}
\end{centering}

The HER is a one-step process  and the reaction path is controlled by the
adsorption energy of hydrogen, whereas the ORR includes three steps with their
associated free energies, $G_0 (U_0)$, $G_1 (U_0)$, and $G_2 (U_0)$. The
corresponding values of the adsorption energies of O and OH  were calculated
previously on the most favorable site of the  thirteen  transition metals
~\cite{MartnezAlonso2021} and they are given in Table \ref{gorr}.  They were
used to determine the free energies for each of the steps and to calculate the
activity according to eq. \eqref{activityorr}, which are also included in Table
\ref{gorr}. The corresponding volcano plot for the catalytic activity as a
function of E$_\mathrm{adsO'}$ is plotted in Figure \ref{volcanoseq}(b). The
highest free energy establishes the rate limiting step for each metal and
determines the catalytic activity. The rate limiting step in most transition
metals is the desorption of HO$^*$ (\ce{HO$^*$ +  H+ + e-  -> H2O + $*$ }) and
the highest free energy is $G_2 (U_0)$. The exceptions to this general
situation are Au, where the rate limiting step is the adsorption of O*
(\ce{1/2O2 + $*$ -> O$^*$}) with the associated free energy  $G_0 (U_0)$, and
Ir and Os, in which the rate limiting step is the formation of HO$^*$
(\ce{O$^*$ + H+ + e- ->HO$^*$}) and the highest free energy is $G_1 (U_0)$. Pt
exhibits the maximum catalytic activity in the volcano plot, followed by Pd,
Ag, and Ir,  in agreement with experimental data ~\cite{Wang2018}. In addition,
these results agree with previous simulations by Norskov {\it et al.}
~\cite{Nrskov2004}, with the only discrepancy of Ir, in which they report that
the rate limiting step corresponds to the desorption of HO. Additionally,  it
is interesting to notice that $G_0 (U_0)$ is very low  for Nb, Mo, W, Re, Fe,
Hf, Ta, V, Cr, and Tc. This corresponds to a very favorable oxygen adsorption,
which could lead to the formation of oxides.

Finally, it should be noted that  metals on the left side of the periodic table
(Hf, Cr, Ta, V, Nb, Mo, and W) with less than half-filled d-bands present the
worst catalytic properties for the HER and the ORR. These results also agree
with previous studies~\cite{Nrskov2005,Quaino2014,Greeley_2006,10.1093/nsr/nwx119}.

\begin{center}
\begin{table}[!]
\caption{Adsorption energy of O ($E_\mathrm{adsO'}$) and OH ($E_\mathrm{adsOH}$) at
equilibrium at the most favorable adsorption site for different transition metals.  The free energies of the three
steps of the dissociative mechanism ($G_0 (U_0)$, $G_1 (U_0)$, and $G_2 (U_0)$) at 300 K and the catalytic activity for  the ORR according to eq. \eqref{activityorr} are also included.}
\hspace*{-1.3cm}
\begin{tabular}{ccccccc}
\toprule
Metal & $E_\mathrm{adsO'}$ (eV) & $E_\mathrm{adsOH}$ (eV) & $G_0 (U_0)$ (eV) & $G_1 (U_0)$ (eV) & $G_2 (U_0)$ (eV) &Activity \\
\midrule
Pt    & 1.66        & 1.19        & -0.79            & 0.29            & 0.49             & -0.23    \\
Au    & 2.43        & 1.52        & 0.06            & -0.02            & -0.04            & -0.69    \\
Cu    & 0.73        & 0.14        & -1.72            & 0.38            & 1.34             & -0.62    \\
Ag    & 2.04        & 0.72        & -0.41            & -0.35           & 0.76             & -0.36    \\
Pd    & 1.25        & 0.85        & -1.20            & 0.57            & 0.63             & -0.29    \\
Ni    & 0.29        & 0.03        & -2.16            & 0.71            & 1.45             & -0.68    \\
Ir    & 0.79        & 0.76        & -1.66            & 0.94            & 0.72             & -0.44    \\
Rh    & 0.45        & 0.27        & -2.00            & 0.78            & 1.21             & -0.56    \\
Cd    & 1.17        & 0.31        & -1.28            & 0.11            & 1.17             & -0.54    \\
Zn    & 0.79        & 0.37        & -1.66            & 0.54            & 1.11             & -0.52    \\
Co    & -0.01       & -0.20       & -2.46            & 0.78            & 1.68             & -0.78    \\
Nb    & -2.17       & -1.80       & -4.62            & 1.34            & 3.28             & -1.53    \\
Mo    & -1.67       & -1.18       & -4.12            & 1.46            & 2.66             & -1.23    \\
W     & -1.65       & -1.01       & -4.10            & 1.61            & 2.49             & -1.16    \\
Re    & -1.19       & -0.64       & -3.64            & 1.52            & 2.12             & -0.98    \\
Ru    & -0.41       & -0.21       & -2.86            & 1.17            & 1.69             & -0.78    \\
Fe    & -0.98       & -1.03       & -3.43            & 0.93            & 2.51             & -1.16    \\
Os    & -0.31       & 0.15        & -2.76            & 1.42            & 1.33             & -0.66    \\
Hf    & -3.67       & -2.54       & -6.12            & 2.10            & 4.02             & -1.87    \\
Ta    & -2.39       & -1.85       & -4.84            & 1.51            & 3.33             & -1.55    \\
V     & -2.99       & -1.54       & -5.44            & 2.41            & 3.02             & -1.41    \\
Cr    & -2.52       & -2.15       & -4.97            & 1.34            & 3.63             & -1.69    \\
Tc    & -1.19       & -0.78       & -3.64            & 1.38            & 2.26             & -1.05   \\
\bottomrule
\label{gorr}
\end{tabular}
\end{table}
\end{center}

\subsection{Volcano plots under mechanical strain}

\begin{figure}[!]
 \centering
  \includegraphics[width=0.8\textwidth]{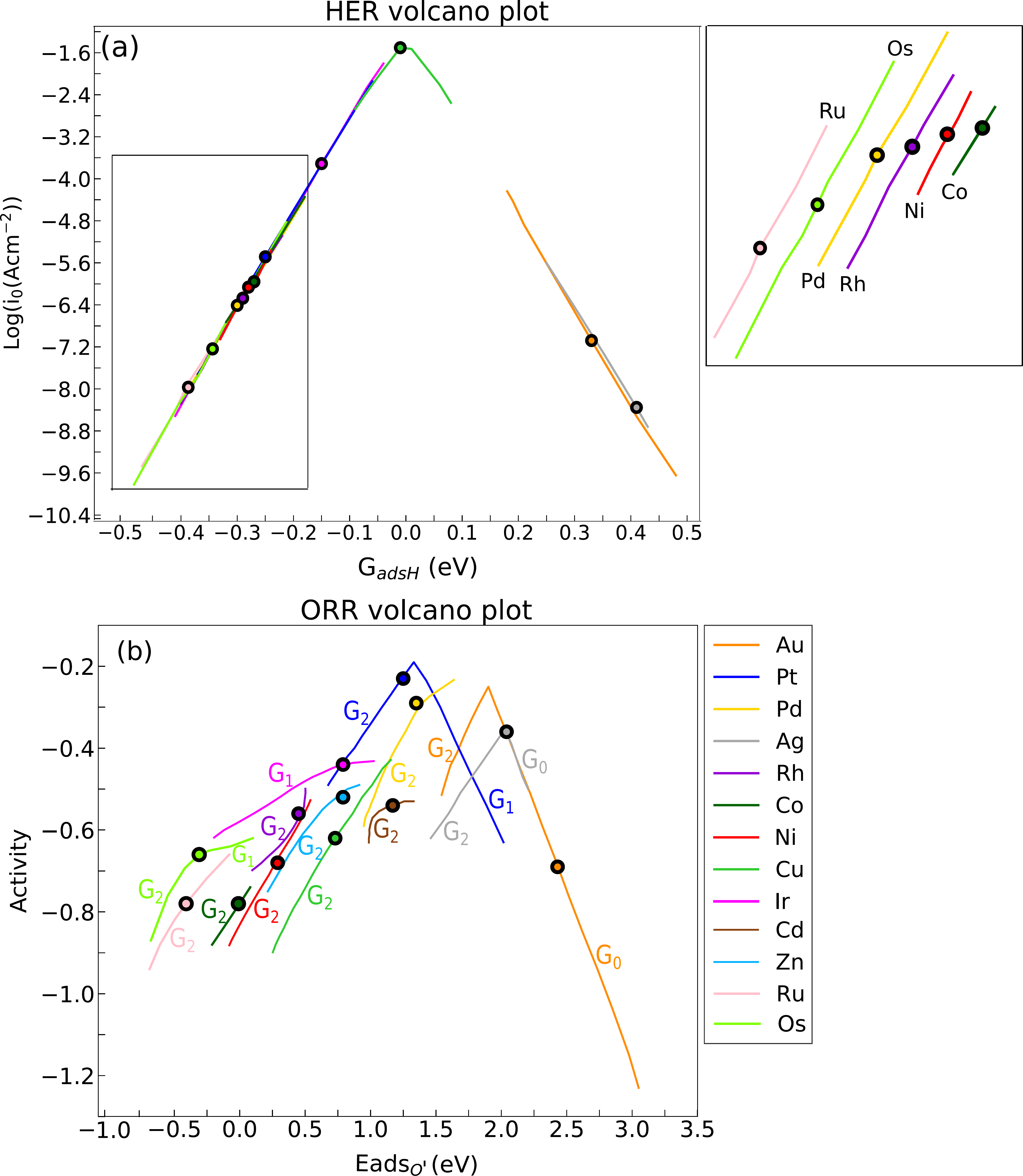}
  \caption{(a) Effect of mechanical strains on the volcano plot for the HER as
a function of $G_\mathrm{adsH}$ for thirteen transition metals. The range of
activities for Pd, Rh, Ni, Co, Ru, and Os  within the rectangle have been plotted to the
right, shifted by 0.05 eV along the horizontal axis for the sake of clarity.
(b) {\it Idem} for the ORR as a function of  E$_\mathrm{adsO'}$. The free
energy that limits the reaction rate is indicated at each branch of the volcano
plot for each metal. The lines stand for the variation in the catalytic
activity of each metal as  a function of the applied elastic strain in the
range from -5\% biaxial compression to 8\% biaxial tension in Ni, Cu, Pd, Ag,
Pt, Au, Rh, Ir, Co, Ru, and Os, and from -2\% biaxial compression to 8\% biaxial
tension in Cd and Zn. Black circles indicate the activity at the mechanical
equilibrium. Cd and Zn are omitted from the HER volcano plot because $\log i_0$
is very low.} 
  \label{volcanosstrain}
\end{figure}

The effect of elastic strains on the catalytic activity of transition metals
for the HER and the ORR is represented by means of volcano plots in Figures
\ref{volcanosstrain}(a) and (b), respectively. These plots were constructed
from the adsorption energies of the biaxially strained slabs of Au, Pt, Pd, Ag,
Rh, Co, Ni, Cu, Ir, Ru, and Os  from -5\% compression to 8\% tension, and of Cd and Zn
from -2\% compression to 8\% tension that can be found in  Figures 6 and 7
of~\cite{MartnezAlonso2021}. These transition metals were chosen due to their
position close to the top of the volcanos in Figure \ref{volcanoseq} and to
their widespread use as heterogeneous catalysts. The black dot in each line  in
Figure \ref{volcanosstrain} stands for the catalytic activity when the applied
strain is zero. The lines for Cd and Zn are not included for the HER reaction
in Figure \ref{volcanosstrain}(a) because they are too far from the top. 

The catalytic activity of the transition metals on the left side of the
volcanos is limited by the desorption of the  products. Compressive strains
increase the adsorption energy of the reactants and avoid the poisoning of the
catalyst surface, improving the catalytic activity. On the contrary, the
catalytic properties of metals on the right side of the volcanos are limited by
the adsorption of the reactants, and can be improved with the application of
tensile strains, which enhance the adsorption of the reactants by reducing the
adsorption energy. In all cases, the effect of elastic strains on the catalytic
activity is limited by the mechanical stability limits of the surface of the
catalyst.

The effect of elastic strains on the catalytic activity (measured by the $\log
i_0$) for the HER  depends on the metal. The largest variations in the
catalytic properties with strain (Figure \ref{volcanosstrain}(a)) are observed
for Ir and Au, where the activity (measured by the $\log i_0$) changes 4.5 and
2.7 units, respectively, when the surface slab is subjected to -5\% biaxial
compression or to 8\% biaxial tension, respectively,  with respect to the
unstrained slab. On the contrary, Co is very insensitive to the mechanical
strains because $E_\mathrm{adsH}$ does not vary with the elastic
strain~\cite{MartnezAlonso2021}.  There are not well established trends for the effect of mechanical strains on the adsorption energies but our previous investigation showed that the influence of the mechanical strain on the adsorption energy increases with the number of electrons in the valence band for the elements in the 4th period of the periodic table (3d). Nevertheless, the effect of mechanical strains on the adsorption energy is much smaller for the 5th and 6th periods ~\cite{MartnezAlonso2021}. 
It should also be noted that the limiting
mechanism (either hydrogen adsorption or desorption) does not change with the
applied strain (the evolution of the catalytic activity with the applied strain
remains to the left or to the right of the volcano cusp) with the exception of
Cu in which $E_\mathrm{adsH}$ is very close to 0 in the unstrained slab. In
this case, tensile or compressive stresses activate one of either rate limiting
steps and reduce the catalytic activity. 
  
The elastic strains also lead to important changes in the catalytic activity of
transition metals for the ORR, as shown in Figure \ref{volcanosstrain}(b). The
largest ones are found in Pt and Au and both metals can be tuned by means of
either compressive or tensile strains, respectively,  to attain the top of the
corresponding volcano plot. This result was already detected for Pt, which is
known to improve its catalytic activity through the application of elastic
strains~\cite{EscuderoEscribano2016,2021} in agreement with our volcano plot.
More interestingly, the application of 2\% biaxial tension can improve the
catalytic activity of Au (which is known to be a bad catalyst for the ORR)
close to that of Pt. In contrast, elastic strains have a minor influence on the
catalytic activity of the ORR for Co and Cd.  The maximum activity of Ir, Rh,
Co, Ni, Cu, Zn, Cd, Pd, Ru, and Os for the ORR is attained when the slab is subjected
to the highest compressive strain, while the best catalytic activity of Ag is
found in the unstrained condition and the application of tensile or compressive
strains always leads to the reduction in activity.

\section{Discussion} 

The results presented above show how the systematic application of elastic
strains can be tailored to optimize the catalytic response of transition metals
for the HER and the ORR. In the former case, the catalytic activity is
determined by $E_\mathrm{adsH}$ which determines the free energy of adsorption,
$G_\mathrm{adsH}$, at a given temperature. The optimum scenario is achieved
when $G_\mathrm{adsH} \rightarrow 0$ through the application of compressive or
tensile strains and the limitations to this strategy are determined by the
sensitivity of $E_\mathrm{adsH}$  to the strain and the maximum strains that
can be applied before mechanical instability limit is achieved, leading to
fracture. The largest changes in catalytic activity with strain for the HER
were found in Pt, Au, and Ir while $E_\mathrm{adsH}$ in Co and Ni was very
insensitive to elastic strains, which did not modify substantially the
catalytic activity. 

It should be noted that the catalytic activity of Cu is overestimated in the
volcano plot for the HER in Figure \ref{volcanoseq}. This discrepancy may be
due to the fact that -- according to equation \eqref{gibbsH} -- the exchange
current density depends strongly on $G_\mathrm{ads_H}$, which in turn is  a
linear function of $E_\mathrm{ads_H}$ (Equation \eqref{gibbsH}). Thus, small
errors in the calculation of $E_\mathrm{ads_H}$ may lead to substantial changes
in catalytic activity, particularly near the cusp of the volcano plot. In order
to test this conjecture, $E_\mathrm{ads_H}$ was calculated again in Pt and Cu
using the metaGGA SCAN functional~\cite{Sun2015} and the methodology presented
in ~\cite{MartnezAlonso2021}. This functional is known to provide more accurate
values of the adsorption energy at a much higher computational cost
\cite{Sun2015}. In the case of Pt, the differences in $E_\mathrm{adsH}$
between the generalized gradient approximation (GGA) with the
Perdew-Burke-Ernzerhof exchange-correlation functional and the metaGGA SCAN
functional were only 0.03 eV but they increased to 0.24 eV in the case of Cu.
Moreover, these differences remained constant as a function of strain  in the case of Pt (supporting information Figure 3). Thus,
the predictions of the adsorption energy for Cu seem to be particularly
sensitive to the selection of the functional and this explains the
overestimation of the Cu activity in the volcano plot of Figure
\ref{volcanoseq}.  And Cu moves to the appropriate position in the volcano plot for the HER in Figure \ref{volcanoseq}(a) when the more accurate adsorption energy (denoted by Cu*) is used to calculate the free energy. 
It should be noted that this sensitivity to the functional could be  less important in
the case of the ORR, because the activity depends on differences in
$E_\mathrm{ads}$ and errors in the calculation of adsorption energies due to
selection of the functional tend to cancel each other. This is evidenced by the
position of Cu in the volcano plot for the ORR in Figure \ref{volcanoseq},
which is in agreement with the experimental data. \ However, further investigation of the variations in the adsorption energy with SCAN in the case of Cu is needed. 

The effect of mechanical strains on the catalytic activity of the ORR is a
little bit more complex because it depends on the free energy of the rate
limiting reaction, which is the maximum of $G_0$, $G_1$, and $G_2$. $G_0$
depends on $E_\mathrm{adsO}$, $G_1$ on $E_\mathrm{adsOH}- E_\mathrm{adsO}$, and
$G_2$ on $-E_\mathrm{adsOH}$, and the influence of the mechanical strains on
the respective adsorption energies is different. 

Tensile strains reduce $E_\mathrm{adsO}$ and $E_\mathrm{adsOH}$ while
compressive strains have the opposite effect. Moreover,  $E_\mathrm{adsO}$ at
zero strain is negative for all the analyzed transition metals while
$E_\mathrm{adsOH}$ is positive under the same conditions for all the analyzed
transition metals (except Co, with $E_\mathrm{adsOH}$ = -0.2 eV). As a result,
the catalytic activity for the ORR of metals in which the rate is controlled by
$G_0$ (adsorption of O$^*$) and $G_2$ (desorption of HO$^*$) are very sensitive
to elastic strains and their catalytic activity increases rapidly with the
application of tensile strains in the former (Au) or compressive strains in the
latter (Cu, Ni, Pt, Pd, Rh, Zn, Cd, Co, Ru, and Os). In some hcp metals, such as Cd and Zn,
the range of elastic strains before mechanical instabilities develop is reduced
and, thus, the tunability of their catalytic activity with mechanical strains
is very limited. Finally, the rate limiting free energy in Ir is $G_1$ (change
from O$^*$ to HO$^*$) that depends on the difference between $E_\mathrm{adsOH}-
E_\mathrm{adsO}$. As both energies evolve in the same direction with strain,
their difference is not very sensitive to the applied strain (Figure
\ref{volcanosstrain}) and Ir activity varies slowly (as compared with other
metals) with deformation. Thus, the efficiency of elastic strains to modify the
catalytic activity not only depends on the catalyst itself but also on the
intermediate species that are present in each reaction step. Some reactions are
more sensitive to the application of elastic strains than others, which opens
the possibility to explore the mechanism to enhance the activity of many other
chemical processes with different intermediate species.

\begin{figure}[t!]
  \centering
  \includegraphics[width=1.0\textwidth]{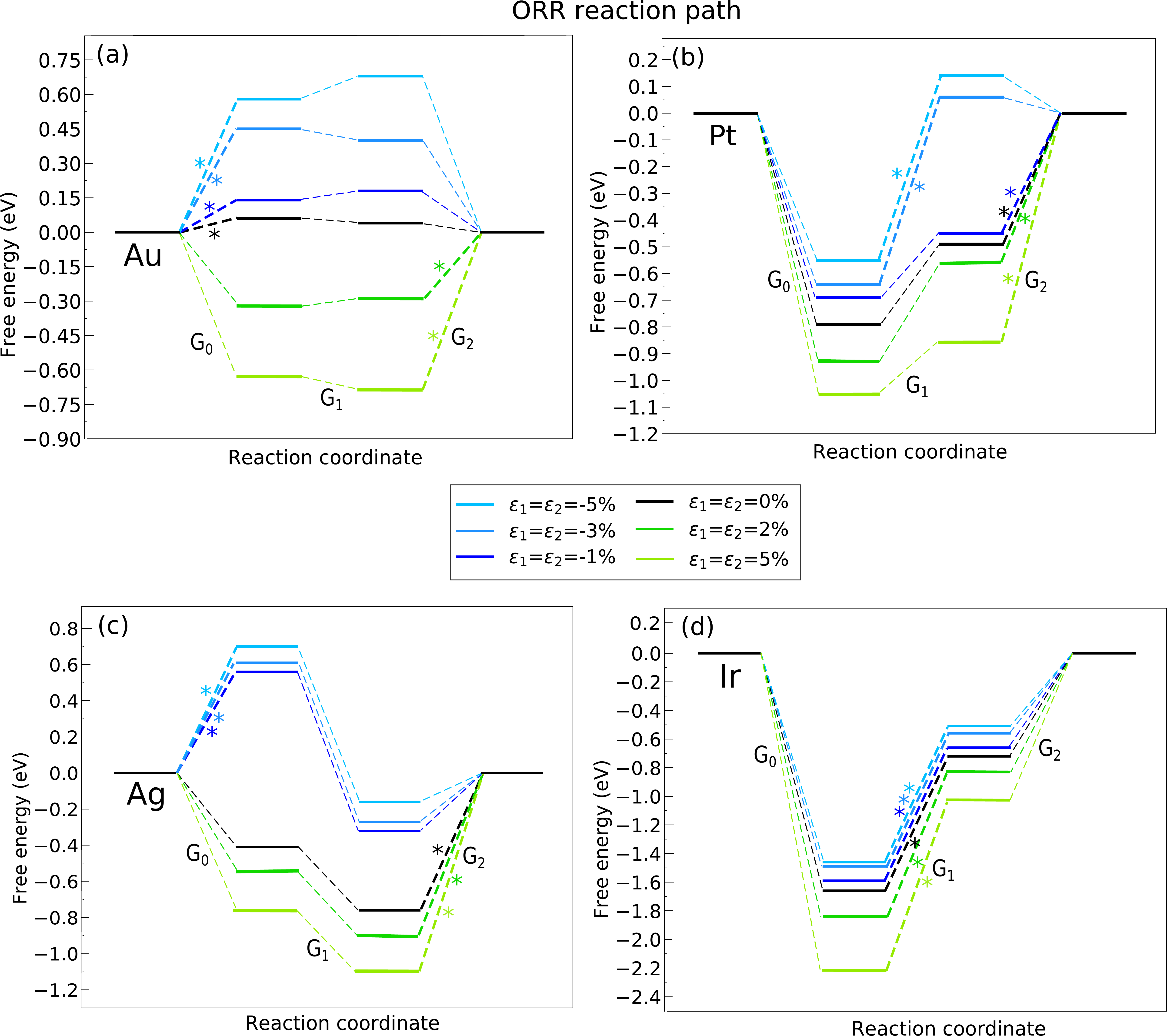}
  \caption{Effect of biaxial elastic strains (in the range -5\% to 5\%) on the free energies along the reaction path for the ORR. (a) Au. (b) Pt. (c) Ag. (d) Ir. The rate limiting step at each strain is indicated with an asterisk.}
  \label{PATHSPASOS}
\end{figure}

In most metals, mechanical strains change the catalytic activity but not the
rate limiting mechanism. This is not the case, however, for Au, Pt, Ag, and Os.
The influence of biaxial strains (in the range -5\% to 8\%) on the free
energies in the ORR reaction path for Au, Pt, and Ag is plotted in Figures
\ref{PATHSPASOS}(a), (b), and (c), respectively. At zero strain, the adsorption
of O is the limiting step in Au (Figure \ref{PATHSPASOS}(a)) and this energy
barrier increases with compressive strains while $G_1$ fluctuates due to the
competition between the adsorption energies of O and HO. As a result,
compressive strains reduce the catalytic activity for the ORR of Au. However,
the application of tensile strains switches the rate limiting step to the
desorption of HO$^*$ because $G_2$ becomes positive and increases rapidly with
the tensile strains while $G_0$ becomes negative and $G_1$ fluctuates between
negative and positive values. The optimum catalytic activity is obtained at
small tensile strains around 2\% biaxial tension.

In the case of Pt (Figure \ref{PATHSPASOS}(b)), the desorption of HO$^*$ is the
limiting step at zero strain, while the adsorption of O is strongly exothermic
and the reaction from O$^*$ to HO$^*$ is slightly endothermic. The application
of compressive strains reduces rapidly the energy barrier for HO$^*$ desorption
and the rate limiting step changes to the reaction from O$^*$ to HO$^*$, which
becomes endothermic. The application of tensile strains increases the energy
barrier for HO$^*$ desorption, leading to a constant reduction in the catalytic
activity. Thus, the maximum catalytic activity is achieved for small
compressive strains (-1.57\% biaxial compression). The rate limiting step also changes from $G_2$ to $G_1$ with the application of compressive strains in the case of Os.

Additionally, unstrained Ag is at the cusp of the volcano plot and the
application of either tensile or compressive strains reduces the activity as
one limiting reaction becomes dominant. In this case, the rate limiting step is
found to be the desorption of HO$^*$ under tensile strains and the O adsorption
under compressive strains (Figure \ref{PATHSPASOS}(c)). Finally, the
application of strain in Ir does not change the rate limiting step of the ORR,
which is the change from O$^*$ to HO$^*$ in all cases (Figure
\ref{PATHSPASOS}(d)). $G_1$ decreases slightly with compressive strains and
increases slightly with tensile strains and the optimum catalytic activity is
found at the maximum compressive strains allowed by the mechanical stability
limits.

\section*{Conclusions}

The effect of elastic strains on the catalytic activity for the HER and the ORR
was analyzed  at (111) surfaces of eight fcc transition metals (Ni, Cu, Pd, Ag,
Pt, Au, Rh, Ir) and (0001) surfaces of five hcp transition metals (Co, Zn,
Cd,  Ru, Os). The catalytic activity for both reactions was determined from the
adsorption  energies for H, O, and OH calculated by density functional theory
under strain. Tensile, compressive, and shear stresses were applied to slabs
until the mechanical stability limit (given by phonon calculations) was
achieved. The volcano plots of the catalytic activity for the HER and the ORR
in the absence of strain were in good agreement with the experimental data in
the literature, validating the theoretical model. It was found that elastic
strains led to significant changes in the catalytic activity of most transition
metals, which could be rationalized as a function of changes in the free energy
of the rate limiting step.

In the case of the HER, it was found that compressive strains increased the
catalytic activity of metals on the ascending branch of the volcanos (Pt, Pd,
Rh, Ni, Ir, Ru, and Os) because they decreased the energy barrier for H$^*$ desorption,
which is the limiting step. On the contrary, tensile strains improved the
activity of metals in the descending branch of the volcano, as they decreased
the energy barrier for H adsorption (Au and Ag). The largest improvements in
activity were found in Au, Ir, and Ag which combined a large sensitivity of the
H adsorption energy  to strain while large mechanical strains could be applied
without leading to failure.

The catalytic activity of the ORR was controlled by the maximum free energy for
the reactions in the dissociative mechanism: adsorption of O, reaction from
from O$^*$ to HO$^*$, and desorption of HO$^*$. In particular, the free
energies associated with the adsorption of O and the desorption of HO$^*$ were
very sensitive to mechanical strains. Thus, the catalytic activity of Au
(controlled by the former) could be enhanced by the application of tensile
strains  while that of Cu, Ni, Pt, Pd, Rh, Zn, Cd, Co,  Ru, and Os (controlled by the
latter) was improved by the application of compressive strains. The optimum
catalytic activity of Ag was found in the unstrained condition and mechanical
deformation always reduced the catalytic activity in this metal. Moreover, only
small elastic strains could be applied to Cd and Zn before the mechanical
instability was reached and elastic strain engineering is not a suitable
strategy to modify the catalytic activity of these metals. Finally, it was also
found that the application of elastic strains could change the rate limiting
step for the ORR in Au, Pt, Ag,  and Os because of the different effect of
mechanical deformation of the free energy of the different intermediate
reactions.

\section*{Acknowledgments}

This investigation was supported by the MAT4.0-CM project funded by the Madrid
region under program S2018/NMT-4381 and by the HexaGB project  (reference
RTI2018-098245) funded by MCIN/AEI/10.13039/501100011033. Computer resources
and technical assistance provided by the Centro de Supercomputaci\'on y
Visualizaci\'on de Madrid (CeSViMa) are gratefully acknowledged. Additionally,
the authors thankfully acknowledge the computer resources at CTE-Power and
Minotauro in the Barcelona Supercomputing Center (project QS-2021-1-0013 and
QHS2021-3-0019).  Finally, use of  the computational resources of the Center
for Nanoscale Materials, an Office of Science user facility, supported by the
U.S. Department of Energy, Office of Science, Office of Basic Energy Sciences,
under Project No. 73377, is gratefully acknowledged. CMA also acknowledges the
support from the Spanish Ministry  of  Education through the Fellowship
FPU19/02031. 

\section*{Conflicts of interest}

There are no conflicts of interest to declare.

\footnotesize


\end{document}